\documentclass[12pt,letterpaper]{JHEP3}
\usepackage{cite}

\title{Bound states and the Bekenstein bound}

\author{
Raphael Bousso\\ 
Center for Theoretical Physics, Department of Physics\\
University of California, Berkeley, CA 94720-7300, U.S.A.\\
{\em and}\\
Lawrence Berkeley National Laboratory, Berkeley, CA 94720-8162, U.S.A.\\ 
E-mail: \email{bousso@physics.harvard.edu}}

\abstract{% 
We explore the validity of the generalized Bekenstein bound, $S\leq\pi
Ma$.  We define the entropy $S$ as the logarithm of the number of
states which have energy eigenvalue below $M$ and are localized to a
flat space region of width $a$.  If boundary conditions that localize
field modes are imposed by fiat, then the bound encounters well-known
difficulties with negative Casimir energy and large species number, as
well as novel problems arising only in the generalized form.  In
realistic systems, however, finite-size effects contribute additional
energy.  We study two different models for estimating such
contributions.  Our analysis suggests that the bound is both valid and
nontrivial if interactions are properly included, so that the entropy
$S$ counts the bound states of interacting fields.}

\preprint{\hepth{0310148} \\ UCB-PTH-03/25}

\begin{document}

\section{Introduction}

Bekenstein's ``universal entropy bound''~\cite{Bek81} states that the
entropy $S$ of any weakly gravitating matter system obeys
\begin{equation}
S\leq 2\pi MR/\hbar
\label{eq-bb}
\end{equation}
where $M$ is the total gravitating mass of the matter and $R$ is the
radius of the smallest sphere capable of enclosing the
system.\footnote{For an antecedent, see Ref.~\cite{Bek74}.  We assume
throughout that the system resides in Minkowski space.  The Bekenstein
bound has sometimes been applied to field systems on a curved
background, and to strongly gravitating systems such as a black hole
or a closed universe.  With few exceptions~\cite{Bou00b}, this is not
sanctioned by its derivation, and we will not concern ourselves with
violations~\cite{KutLar00} arising in this way.---We set $\hbar=k_{\rm
B}=c=1$, except in the introduction, where $\hbar$ is displayed
explicitly.  Newton's constant, $G$, is always displayed.}

Originally this bound arose from a gedankenexperiment by which the
matter system is dropped into a large black hole.  Classical general
relativity implies that the black hole horizon area increases by
$\Delta A\leq8\pi GMR$ in this process.  By the generalized second law
of thermodynamics~\cite{Bek72,Bek73,Bek74} (GSL), the total entropy
visible to an outside observer will not decrease; hence, $S\leq\Delta
A/4G\hbar$ and Eq.~(\ref{eq-bb}) follows (see, e.g., Ref.~\cite{Bou02}
for a review).  We will not address here the controversial question
concerning the extent to which quantum effects modify or invalidate
this derivation from the
GSL~\cite{UnrWal82,UnrWal83,PelWal99,Bek99,MarSor02} (for reviews, see
Refs.~\cite{Wal99,Bek01b}).

The present paper studies aspects of a different question: Does the
Bekenstein bound actually hold in nature?  We are motivated to revisit
this issue by a recent re-derivation~\cite{Bou03} of Bekenstein's
bound from the generalized covariant entropy bound~\cite{FMW} (GCEB),
from which Eq.~(\ref{eq-bb}) emerges in a somewhat stronger form:
\begin{equation}
S\leq \pi M a/\hbar.
\label{eq-gbb}
\end{equation}
Here $a$ is the smallest distance between any two parallel planes that
bracket the system; a more precise definition is given in
Ref.~\cite{Bou03}.  Note that $a\leq 2R$, so that the ``generalized
Bekenstein bound'', Eq.~(\ref{eq-gbb}), implies the original
Bekenstein bound, Eq.~(\ref{eq-bb}).  (For example, for a rectangular
box, $a$ is the length of the shortest edge.  For a sphere, $a=2R$,
and the two expressions agree.)  Unlike the derivation of the
Bekenstein bound from the GSL, the new derivation from the GCEB takes
place in flat space, which excludes corrections from Unruh radiation.

The GCEB is a stronger version of the covariant entropy
bound~\cite{ceb1} (CEB), a conjecture concerning the entropy of matter
in arbitrary spacetime regions.  Compared to the GSL, the covariant
bounds are less well-established hypotheses.  However, they are more
general (for example, they imply the GSL as a special case), and they
are essential for putting the holographic
principle~\cite{Tho93,Sus95,ceb2} on a firm footing.  If true, the
covariant bounds will thus be of wider significance in quantum
gravity.

This tempts us to regard the GCEB as the more fundamental conjecture,
and to consider the Bekenstein bound a consequence of the GCEB.  Tests
of the GCEB generally require the computation of families of null
geodesics, which can be complicated.  The Bekenstein
bound\footnote{From here on, the term will refer to the generalized
Bekenstein bound, Eq.~(\ref{eq-gbb}), unless specifically stated
otherwise.}  is more easily applied to examples, because the mass and
size of many systems are known or easily determined.  Thus the
Bekenstein bound offers a simple way of testing the GCEB.

In Refs.~\cite{FMW,BouFla03}, sufficient phenomenological conditions
were identified under which the GCEB (and, directly or by extension,
the Bekenstein bound) hold automatically.  In order to determine its
wider validity, our study of the Bekenstein bound will focus chiefly
on examples in which the assumptions of Refs.~\cite{FMW,BouFla03} do
not hold.  Roughly speaking, this requires consideration of systems
whose entropy is dominated by modes whose wavelength is of order the
largest dimension.

To test Eq.~(\ref{eq-gbb}), one needs precise definitions of $S$, $M$,
and $a$.  Which type of entropy (canonical, microcanonical,
entanglement, etc.) does the bound refer to?  Different choices may
agree in the high-temperature regime, but they are certainly
inequivalent in those systems which are most likely to violate the
Bekenstein bound.\footnote{The same systems are trivial in the context
of the original covariant entropy bound~\cite{ceb1}, which is thus
less sensitive to the precise definition of $S$ than the GCEB.  Indeed
several proposals of cosmological entropy bounds were falsified by
simple examples whose entropy is unambiguous (see, e.g.,
Refs.~\cite{FisSus98,KalLin99,Bou02}).}  What definition of energy
should be used: the energy above the ground state, or the gravitating
mass; an eigenvalue or an expectation value?  Geometrically, the width
$a$ is sharply defined in the weak gravity limit. But what is the
radius of a quantum state?  Not surprisingly, analyses using different
definitions come to different conclusions about the validity of the
Bekenstein bound~\cite{Bek81,UnrWal82,Unw82,Pag82,%
Deu82,Bek82,Bek83,Bek84,%
UnrWal83,SchBek89,SchBek90,BekSch90,Bek94,Bek94b,Bek99,%
Pag00a,Pag00b,Pag00c,Bek00b}.

Our point of view is agnostic---we do not claim to know the correct
formulation of Bekenstein's bound {\em a priori}.  Rather, we would be
content to find any reasonable definition of $S$, $M$, and $a$ such
that the Bekenstein bound becomes a precise, non-trivial, and
empirically true statement.  In the spirit of the covariant entropy
bound, the general goal is to identify laws governing the information
content of spacetime regions---patterns whose origin and implications
may be of some significance.

Here we examine the validity of the Bekenstein bound with a
microcanonical definition of entropy:
\begin{equation}
S\equiv\log {\cal N}(M),
\end{equation}
where ${\cal N}(M)$ the number of energy eigenstates with eigenvalue
$E\leq M$.  This definition is presented in detail in
Sec.~\ref{sec-define}.  It is motivated empirically by additional
difficulties that arise if canonical ensembles are
considered~\cite{Deu82}, or if states other than energy eigenstates
are admitted~\cite{Pag00c,BouFla04}.

We must also ensure that the systems we consider have finite width
$a$.  Initially, we enforce this by decree: we impose rigid boundary
conditions restricting field modes to finite regions.  We study mainly
systems of scalar fields in finite cavities.  In Sec.~\ref{sec-ok}, we
verify that the bound is easily satisfied at high temperatures, where
the canonical and microcanonical ensembles agree.  We note that the
bound becomes nearly saturated when the typical energy of quanta drops
to the inverse size of the system.

In Sec.~\ref{sec-bad}, we construct situations in which the Bekenstein
bound is apparently exceeded.  Field systems with negative Casimir
energy seem to violate Eq.~(\ref{eq-bb}) by permitting its right hand
side to become negative.  Moreover, by proliferating the number of
field species, the entropy of a cavity of fixed size and fixed energy
can apparently be increased above (\ref{eq-bb}).  In addition to these
two well-known difficulties we illustrate a new problem that occurs
only in the tighter version (\ref{eq-gbb}) of the bound: Because the
smallest width enters the right hand side, the entropy can apparently
be made arbitrarily large by increasing the transverse area of a thin
system at fixed energy and width (Sec.~\ref{sec-transverse}).

In response to arguments raised against his bound,
Bekenstein~\cite{Bek82,Bek83,Bek00b} has insisted that
Eq.~(\ref{eq-bb}) applies only to complete systems.  Indeed, each of
the above examples is an incomplete system, because fields were
assumed to be localized to finite regions, but the matter causing this
restriction was not included in the total energy.

Bekenstein has treated such examples by identifying missing parts and
estimating their minimum mass~\cite{Bek82,Bek83,Bek00b} (and in some
cases, also the additional entropy).  This is satisfactory if the goal
is to dispute that a given setup violates the bound.  However, the
procedure has been somewhat ad hoc, varying strongly depending on the
specific example considered.  It has not led to a clear definition of
what does, in fact, constitute a ``complete system'', whose entropy
and mass could be fully calculated in a single step.

In order to find an appropriate criterion, we will try to identify a
single source of extra energy which contributes sufficiently to each
and every one of the problematic examples so as to validate the bound.
In Secs.~\ref{sec-wall} and~\ref{sec-pressure}, we devise two
different models for completing the partial systems of
Secs.~\ref{sec-ok} and~\ref{sec-bad}.  Each model implements one
particular aspect that might be considered necessary for the spatial
restriction of fields: confinement by interactions, and stability
against pressure.  We evaluate both models according to their ability
to restore the validity of the Bekenstein bound without rendering it
trivial.

In Sec.~\ref{sec-wall}, we note that the energy eigenstates of free
fields are delocalized; to restrict to finite spatial width,
interactions are essential.  We model the confinement of the active
fields to a region of with $a$ by couplings to a stationary background
potential.  We estimate that the gradient energy of the background
gives a lower bound on the energy we had failed to include in our
previous analysis.  We find that that this extra contribution was
rightfully neglected in the unproblematic examples of
Sec.~\ref{sec-ok}.  However, in the apparent counterexamples of
Sec.~\ref{sec-bad}, confinement effects are dominant.  They
significantly increase the total energy and restore the validity of
Bekenstein's bound.

In Sec.~\ref{sec-pressure}, we note that field modes restricted to a
finite region exert pressure (or suction) on the region's
boundary~\cite{Bek82}.  This pressure must be balanced by a tensile
wall or ``container'', whose minimum energy we calculate as a function
of its shape and the orthogonal pressure.  This generalizes earlier
estimates~\cite{Bek82} to nonspherical systems.  We find that this
model is not satisfactory.  Namely, the containment contribution by
itself is not sufficient to eliminate counterexamples, although it
does give large corrections where they are not needed (in the examples
of Sec.~\ref{sec-ok}).  Moreover, we argue that a container, and the
resulting extra contribution to the energy, are in principle
avoidable.

In Sec.~\ref{sec-conclusions}, we conclude that confinement by
interactions constitutes an important physical mechanism by which the
restriction to finite spatial width contributes to the total energy of
a system, whereas the container model does not appear to capture an
essential requirement.  Our analysis shows that interactions play a
crucial role and should be properly included from the start.  Applying
the bound to free fields obeying external boundary conditions robs the
bound of its physical content.  One must consider real interacting
theories, which contain bound states of finite spatial extent.  The
entropy $S$ is thus the logarithm of the number of bound states whose
spatial width does not exceed $a$ and whose energy eigenvalues satisfy
$E\leq M$.

\section{Modes, states, and entropy}
\label{sec-define}

In this section we describe in detail how the Bekenstein bound is
applied to field systems confined to compact spatial regions.  This
involves the construction of a Fock space and the counting of
admissible energy eigenstates.  As a starting point, we enforce field
localization by imposing external boundary conditions---an approach we
will dismiss as unphysical in Sec.~\ref{sec-wall}.

Consider a free scalar field $\phi(t,x,y,z)$, of mass $\mu$, occupying
a stationary cavity of arbitrary size and shape.  We will assume that
$\phi$ respects Dirichlet boundary conditions ($\phi=0$) on all cavity
walls.  The Fock space construction begins by expanding $\phi$ into
energy eigenmodes,
\begin{equation}
\phi(x,y,z,t) = \sum_{\mathbf k} a_{\mathbf k}\, 
u_{\mathbf k}(x,y,z)\, e^{i \omega t} + \mbox{c.c.},
\end{equation}
whose spatial factor obeys the time-independent Schr\"odinger equation
\begin{equation}
\nabla_i \nabla^i u_{\mathbf k} = (\omega^2-\mu^2) u_{\mathbf k},
\end{equation}  
subject to the assumed Dirichlet boundary conditions.

Acting on the vacuum, the operator $a_{\mathbf k}^\dagger$ creates a
one particle state of energy $\omega(\mathbf k)$.  More general
energy eigenstates of the Fock space are constructed by acting with a
variety of creation operators $a_{{\mathbf k}_1}$, $a_{{\mathbf
k}_2}$, etc., one or multiple times.  Thus, each energy eigenstate is
labeled by a set of occupation numbers $\{N_{\mathbf k}\}$ listing the
number of particles in each mode ${\mathbf k}$.  The corresponding
energy eigenvalues are
\begin{equation}
E_\phi(\{N_{\mathbf
 k}\}) = \sum_{\mathbf k} N_{\mathbf k} \omega(\mathbf k). 
\end{equation}
These states form a complete orthonormal basis of the Fock space.

We will generally consider a Lagrangian with $Q$ such scalars,%
\footnote{At high temperature we may think of $Q$ more
generally as an {\em effective} number of massless scalars,
representing also massive scalars, as well as fermions and vector
fields whose mass is negligible compared to the characteristic
temperature.  For example, in a cavity with conducting walls, for
temperatures well below .5 MeV, we have $Q=2$, corresponding to the
polarization states of the photon.}
$\phi_q$, $1\leq q\leq Q$, which will be assumed not to interact
mutually.  Then eigenstates are characterized by occupation numbers
$\{N^{(q)}_{\mathbf k}\}$ and energy eigenvalues
\begin{equation}
E_\phi(\{N^{(q)}_{\mathbf k}\}) = 
\sum_{{\mathbf k}, q} N^{(q)}_{\mathbf k} \omega^{(q)}(\mathbf k). 
\end{equation}

We define the entropy $S(M)$ microcanonically:
\begin{equation}
S(M) = \log {\cal N}(M),
\label{eq-sm}
\end{equation}
where ${\cal N}(M)$ is the number of energy eigenstates with energy
$E\leq M$.  This is equivalent to the statement that $S=\log\dim{\cal
H}(M)$, where ${\cal H}(M)$ is the Hilbert space spanned by the energy
eigenstates with $E\leq M$.  Though we obtain $S$ by counting pure
states, note that a definition of $S$ in terms of a density matrix
$\hat\rho$ would be equivalent, as long as $\hat\rho$ is constructed
entirely from the Hilbert space ${\cal H}(M)$.  This is because
$\mbox{Tr~}\hat\rho\log\hat\rho\leq\log{\cal N}(M)$, with equality
holding for the density matrix with equal weights for all states.
However, definitions of $S$ that permit the admixture of states with
$E>M$ by superposition or in a density matrix are inequivalent and
will be explored elsewhere~\cite{BouFla04}.  In the present paper, $M$
will be regarded as a sharp cutoff for energy eigenstates, not just as
a cutoff for energy expectation values.

Other choices would have been conceivable, for example a canonical
definition.\footnote{This would define $S$ as the entropy of the
density matrix of states at the temperature at which the energy
expectation value is $M$.  With this definition, apparent violations
of Bekenstein's bound, in addition to those discussed in
Sec.~\ref{sec-bad}, would occur at very low
temperatures~\cite{Deu82}.  However, those violations would be
resolved by the effects discussed in
Sec.~\ref{sec-wall}~\cite{Bek00b}.}  However, it is already of
considerable interest to show that there exists {\em some\/} set of
definitions under which Bekenstein's bound is well-defined,
non-trivial, and not violated.  Our choice is guided by the conceptual
simplicity of counting pure states, and by the fact that Bekenstein's
bound contains $M$ (so it is natural to regard energy, not
temperature, as the relevant macroscopic quantity to be fixed, and to
work with energy eigenstates).

The Bekenstein bound does not contain Newton's constant, and it can be
tested entirely within the non-gravitational setting of quantum field
theory in flat space.  However, in that context the bound might appear
to be meaningless because there is no unique notion of absolute
energy---one can always shift all values by a constant
amount~\cite{Deu82}.  In order to define $E$ we must recall the
derivation of the Bekenstein bound from the focussing of
light-rays~\cite{Bou03} (before the zero gravity limit is taken): $E$
is the gravitating mass of the matter system.

A central issue to be debated in this paper is the question of how $E$
should be calculated.  We will begin by defining $E$ as the total
energy of the $\phi$ field(s) in the cavity:
\begin{equation}
E=E_\phi + E_{\rm C}.
\label{eq-e0}
\end{equation}
Here $E_{\rm C}$ is the Casimir energy, i.e., the energy of the vacuum
state, which may be non-zero due to quantum effects.  At a later stage
(Sec.~\ref{sec-wall}), we will refine this definition of $E$ and
include additional contributions which we neglect for now.

The larger the field mass $\mu$, the fewer states will be allowed
below a fixed energy $M$, because some of the energy allotment is
diverted into rest mass, which carries no entropy.  Since we are
interested in challenging Bekenstein's bound, we shall set $\mu=0$ and
consider only massless fields from here on.

\section{Scalar fields in a cavity}
\label{sec-ok}

In this section we verify that the Bekenstein bound,
Eq.~(\ref{eq-gbb}), is satisfied by systems at high temperature, and
that it becomes nearly saturated in the regime where the thermodynamic
approximations begin to break down.

\subsection{High temperature limit}
\label{sec-thermal}

In general, the microcanonical entropy, Eq.~(\ref{eq-sm}), is
difficult to calculate exactly.  However, at high temperatures it
agrees with the entropy of a corresponding canonical ensemble, which
can be easily computed in a thermodynamic approximation.\footnote{In
this regime, the assumptions of Ref.~\cite{BouFla03} can be shown to
be satisfied.  Thus, we could simply appeal to that analysis.  In the
present context, however, explicit verification of the bound is
straightforward and more instructive.}  For the thermodynamic
description to be valid, the discreteness of the energy spectrum due
to the finite size of the cavity should not be noticable.  Thus we
require that the characteristic wavelength of the radiation, $1/T$, be
short compared to the length scales of the cavity.  In particular,
\begin{equation}
T\gg \frac{1}{a}.
\label{eq-lx}
\end{equation}
For the same reason, the temperature should also be much larger than
the mass of the scalar field.  This is not an important restriction:
in any case we generally have massless fields in mind since they allow
for the most entropy at a given energy.

Consider the canonical ensemble of the field(s) $\phi$ at the
temperature $T$.  With other parameters fixed, the energy expectation
value, $\langle E\rangle(T)$, is a monotonic function of the
temperature; we denote the inverse function by $T(E)$.  Let ${\cal
S}(T)$ be the entropy in the canonical ensemble.  Then the
microcanonical entropy of Eq.~(\ref{eq-sm}) is given by
\begin{equation}
S(M) \approx {\cal S}[T(M)]
\label{eq-scsm}
\end{equation}
to an excellent approximation.

The expectation value for the total energy of the scalar field(s) is
\begin{equation}
\langle E\rangle(T) = {\pi^2\over 30} Q V T^4,
\label{eq-et}
\end{equation}
where $V$ is the volume of the cavity.  Recall that $Q$ is the
effective number of scalar degrees of freedom.  The canonical entropy
is
\begin{equation}
{\cal S}(T) = {4\over 3} \left({\pi^2\over 30}\right) Q V T^3.
\label{eq-st}
\end{equation}
By inverting Eq.~(\ref{eq-et}) and using Eq.~(\ref{eq-scsm}) we find
that the microcanonical entropy obeys
\begin{equation}
{S\over\pi Ma} \approx {4\over 3\pi Ta}\ll 1.
\end{equation}
in the high temperature regime.  Since $Ta\gg 1$ by assumption, the
bound (\ref{eq-gbb}) holds comfortably.  Except for this condition,
the result is independent of the shape and volume of the cavity.

Note that our result is also completely independent of the number of
species, $Q$.  This may seem surprising.  After all, as $Q$ grows,
there are more ways to distribute the total energy into combinations
of particles, so the entropy should increase without bound as
$Q\to\infty$.  However, this argument applies only if the total
energy, or $\langle E\rangle$, is fixed.  Here, we are holding the
temperature at a fixed value, which must obey Eq.~(\ref{eq-lx}).  By
Eq.~(\ref{eq-et}), increasing $Q$ at fixed $T$ leads to an increase in
$\langle E\rangle$.  Increasing $Q$ at fixed $\langle E\rangle$,
however, is accompanied by a decrease in $T$.  The above calculation
shows that violations of the bound cannot occur before $T$ drops below
the minimum value required for the validity of the canonical
approximation, set by Eq.~(\ref{eq-lx}).

\subsection{Saturation limit}
\label{sec-cube}

Having verified that the bound is always satisfied at temperatures
much larger than the inverse size of a system, let us approach the
opposite limit in which the thermodynamic approximation breaks down.
If we extrapolate the high temperature expressions, (\ref{eq-et}) and
(\ref{eq-st}), to the temperature $T\approx 1/a$, we find that the
entropy and the right hand side of the bound are both of order $Q$,
suggesting that the bound becomes saturated and perhaps even violated.

To resolve this issue, we can no longer rely on the canonical
approximation to the microcanonical entropy.  The discrepancy with the
microcanonical entropy becomes a factor of order unity when $Ta\approx
1$.  For energies around and below $Q/a$, the microcanonical entropy
must be evaluated by explicit state counting.  Before considering the
extreme deformations (very large species number, or very thin cavity
shapes) that lead to apparent difficulties, let us verify the validity
of the bound for the generic case of a few species confined to a
cavity whose various dimensions are roughly equal.

At low energy, the microcanonical entropy depends sensitively on the
field content and the shape of the cavity, and on the boundary
conditions employed.  This forces us to select a specific (though
representative) example, say a single ($Q=1$) massless scalar field
confined to a cubic cavity of side length $a$, with Dirichlet boundary
conditions.  This example, and many others, have been studied (in the
context of the original Bekenstein bound) in Ref.~\cite{Bek84}.  We
reproduce here only the steps necessary to demonstrate explicitly how
the entropy $S(M)$ is calculated and quote the main results.

The mode solutions are
\begin{equation}
\phi_{klm} = (2\omega)^{-1/2} \left(\frac{2}{a}\right)^{3/2}
\sin {k\pi x\over a} \sin {l\pi y\over a} \sin {m\pi z\over a}
e^{i\omega t},
\end{equation}
with frequency
\begin{equation}
\omega =  {\pi\over a}\left( k^2+l^2+m^2 \right)^{1/2},
\label{eq-o1}
\end{equation}
where $k$, $l$, and $m$ run over the positive integers.  We wish to
verify that
\begin{equation}
S(M)\equiv \log {\cal N}(M)\leq \pi Ma
\end{equation}
for all values of $M$.  We will set $E_{\rm C}=0$ until
Sec.~\ref{sec-casimir}, where some implications of nonzero
Casimir energy will be considered.

In calculations both of the entropy, and of the bound, only the
product $Ma$ enters; the overall scale of the system is not relevant.
In the cubic example considered here, the smallest light-sheet width
is $a$.  In order to avoid cluttering our expressions with factors of
$\pi/a$ from Eq.~(\ref{eq-o1}), we set
\begin{equation}
a=\pi.
\end{equation}
Expressions for general $a$ can be recovered by multiplying every
energy by a factor $a/\pi$.

The only state with $0\leq M<\sqrt{3}$ is the vacuum, so $S=0$ and the
bound is obeyed.  For $M=\sqrt{3}$, a new state is allowed, namely
$k=l=m=1$ with occupation number $1$; hence, ${\cal N}(\sqrt{3})=2$.
Again the bound holds, since $S/\pi Ma = \log 2 /\sqrt{3}\pi^2 < 1$.
For $M=\sqrt{6}$, there are 3 additional states ($k=l=1$, $m=2$ and
permutations, with occupation number one), so that ${\cal N}=5$.
However, $\log 5/\sqrt{6}\pi^2 \approx \frac{1}{11}$ is still less
than unity.  For larger $M$, one finds that the entropy increases to
within an order of magnitude of saturating the bound, before
asymptoting to the thermodynamic behavior
\begin{equation}
{S\over \pi^2 M} \to {4\pi\over 3} \left({\pi\over 30M}\right)^{1/4}.
\end{equation}
More details are found in Ref.~\cite{Bek84}; see Fig.~1 therein.

To summarize, for a single scalar field ($Q=1$) in a cube, the bound
(\ref{eq-gbb}) is obeyed for any value of the total energy.  The
closest approach to saturation occurs at small values of $M$ ($\approx
10/a$); at larger values, explicit state counting exhibits the onset
of the asymptotic thermodynamic behavior $S\sim M^{3/4}$ described in
Sec.~\ref{sec-thermal}.

\section{Challenges to the bound}
\label{sec-bad}

In this section we describe various violations of the bound
(\ref{eq-gbb}) which may occur with the definition (\ref{eq-e0}) of
the energy.  In later sections we will argue that there are additional
contributions to the energy which can salvage the bound.

\subsection{Casimir problem}
\label{sec-casimir}

The finite size of the cavity gives rise to a shift of the vacuum
energy density in its interior, whose integral is the Casimir energy,
$E_C$.  So far we have neglected this contribution.  However, the
Casimir energy is an inseparable part of the total gravitating energy
$E$ of any confined field, and our definition (\ref{eq-e0}) calls for
its inclusion.  (There are further contributions arising from the
enforcement of boundary conditions, which are not yet included in $E$
and will be treated in later sections.)  The Casimir energy, including
its sign, depends sensitively on the field type, the detailed shape of
the cavity, and the type of boundary conditions\cite{BorMoh01}.

There are many well-known examples in which the Casimir energy is
negative.  For such cases it is immediately obvious that the bound can
be violated.  Simply choose $M$ negative but of smaller magnitude than
$E_{\rm C}$.  Then there is at least one state (the vacuum) with
energy below $M$, and we have $S\geq 0$.  Thus the entropy exceeds the
bound, $\pi M a$, which is negative.  Note that this difficulty occurs
also in the original Bekenstein bound, Eq.~(\ref{eq-bb}), in which the
largest dimension of the system is used.

Though the above argument suffices to pose the Casimir problem in
principle, a slightly more quantitative statement will later be needed
to determine if we are able to resolve the problem.  The Casimir
energy is not generally known in closed form, but it can be calculated
for many special cases and limits, including the rectangular cavities
studied here.  It will be sufficient to note that the energy of the
vacuum state, in known examples, receives a Casimir correction which
is of the general form
\begin{equation}
E_{\rm C} = \sum_{i=1}^Q {\eta_i A \over a^3}.
\label{eq-ecest}
\end{equation}
Here $a$, the width of the system, is assumed to be the smallest
dimension; $A$ is the transverse area.  $Q$ is the effective number of
scalar field degrees of freedom.  The numbers $\eta_i$ depend on the
field type and the exact shape of the cavity.  In known examples, the
$\eta_i$ are typically of order $10^{-1}$ or smaller and need not be
positive~\cite{BorMoh01}.  Physically, this expression arises because
the Casimir energy density is set by the smallest dimension of the
system and hence is of the form $\eta_i/a^4$ for each species.
Integration over the volume and summation over species yields
Eq.~(\ref{eq-ecest}).  This general estimate reduces to the results of
Ref.~\cite{Bek00b} for the case of an electromagnetic field confined
to certain cavities.
%Lukosz: $E_{\rm C}$ is approximately $1/11a$, $-c/26.2 a^2$,
%and $-bc/73a^3$, respectively.

\subsection{Species problem}
\label{sec-species}

If the number of sufficiently light field species in a cavity, $Q$, is
increased, the entropy will grow even though the total energy is held
fixed.  This is because for each particle in each state, one has an
additional choice of species.  Hence, the number of states with
$N=\sum_{qklm} N^{(q)}_{klm}$ particles and energy $E_\phi\leq M$
increases by a factor of $Q^N$, while the right hand side of the bound
remains unchanged.  Thus, for any $M$ large enough to allow at least
the $Q$ different one-particle states, there always exists some
critical species number $Q_{\rm vio}$ such that the bound is violated
for all $Q\geq Q_{\rm vio}$~\cite{Pag82}.  This is known as the
``species problem''.\footnote{There are variants of this problem, for
example a single species with extremely large spin.  The analysis and
resolution proceeds along similar lines.}

The species problem is most acute when the number of quanta is
small.  Consider, for example, the cubic cavity introduced above and
choose $M$ so that at most a single particle of the lowest energy is
admitted: $M=\sqrt{3}$.  With $Q=1$, we found in the previous section
that $S=\log 2$, whereas the right hand side was somewhat larger,
$\sqrt{3}\pi^2$.  With $Q$ noninteracting scalars, however, there are
$Q$ orthogonal one-particle states, plus the vacuum, all of which have
$E\leq M$.  Thus the entropy is given by
\begin{equation}
S=\log (1+Q)
\end{equation}
In this example the bound will be violated for $Q\geq
e^{\sqrt{3}\pi^2}-1\approx 3\times 10^7$.  

Of course, there is no evidence that the actual number of species in
Nature is so large; in this sense, the species problem is not manifest
empirically.  Indeed, one could regard the Bekenstein bound as a
prediction of an upper limit on the number of field degrees of
freedom~\cite{Bek99}.  We do not adopt this point of view here,
because the model of Sec.~\ref{sec-wall} suggests that when all
contributions to the mass are included, the bound actually becomes
more and more easily satisfied at large species number.

Note that the species problem does {\em not\/} occur in the covariant
bound, since the length scale at which semi-classical gravity breaks
down due to one-loop effects is $\ell_{\rm Pl}\sqrt{Q}$.  The
Bekenstein bound, however, does not contain Newton's constant.
Arguments involving renormalization of $G$~\cite{SusUgl94} cannot be
advanced in its defense.

\subsection{Transverse problem}
\label{sec-transverse}

Finally, we consider thin cavities, which give rise to a problem that
is unique to the generalized Bekenstein bound (\ref{eq-gbb}), as
opposed to Eq.~(\ref{eq-bb}).  A thin system is one with very unequal
dimensions.  The width, $a$, can be chosen to coincide with the
smallest dimension~\cite{Bou03}.  A large transverse size allows
orthogonal modes with nearly equal energy, which differ only in their
large transverse wavelength.  This gives rise to high entropy at low
energy.  Hence, for thin systems, the bound can be violated even for
$Q=1$.  We call this the transverse problem.

There are two limits of interest, roughly the shapes of a pencil and
of a pancake.  We will consider these limits for a rectangular cavity
of side lengths $a$, $b$, and $c$.  Without loss of generality, we
assume that
\begin{equation}
a\leq b\leq c.
\end{equation}
(The special case $a=b=c$ was already considered above.)  With
Dirichlet boundary conditions, the mode solutions for a general
rectangular cavity are
\begin{equation}
\phi_{klm} = \frac{2}{\sqrt{\omega abc}}
\sin {k\pi x\over a} \sin {l\pi y\over b} \sin {m\pi z\over c}
e^{i\omega t},
\label{eq-rmds}
\end{equation}
with frequency
\begin{equation}
\omega = \pi \left( {k^2\over a^2}+{l^2\over b^2}+{m^2\over c^2}
\right)^{1/2},
\label{eq-omegas}
\end{equation}
where $k$, $l$, $m$ are positive integers.

\subsubsection{Pencil-shaped cavity}
\label{sec-pencil}

First, consider an elongated cavity with $a=b\ll c$.  Instead of
$c$, we find it convenient to use the dimensionless aspect ratio
\begin{equation}
\epsilon \equiv \frac{a}{c} \ll 1.
\end{equation}
We choose the light-sheet that gives the shortest width, $a$, and we
fix the overall scale of the cavity by choosing
\begin{equation}
a=\pi
\end{equation}
as in the cubic example.  Equation (\ref{eq-omegas}) simplifies to
\begin{equation}
\omega = \left( k^2+l^2+m^2 \epsilon^2 \right)^{1/2}.
\end{equation}

For $0 \leq M < \sqrt{2+\epsilon^2}$, the bound is obeyed as the
vacuum is the only state.  For larger $M$, more and more single
particle states with $k=l=1$ and increasing $m$ are admitted.  In this
mass range,
\begin{equation}
\sqrt{2+\epsilon^2}\leq M < \sqrt{5+\epsilon^2},
\label{eq-pencilrange}
\end{equation}
the number of states is given exactly by
\begin{equation}
{\cal N}(M) = 1+ \frac{\sqrt{M^2 - 2}}{\epsilon},
\label{eq-nmpencil}
\end{equation}
where it is understood that the integer part is taken.

Eq.~(\ref{eq-nmpencil}) shows that for $M>\sqrt{2}$, the density of
states can be made arbitrarily large by choosing the edge $c$ to be
much longer than $a$ ($\epsilon\ll 1$).  The entropy grows
logarithmically with the transverse area $A$:
\begin{equation}
S \approx -\log \epsilon = \log \frac{A}{a^2} .
\label{eq-spencil}
\end{equation}
Thus, $S$ can be increased arbitrarily while keeping $M$ and $a$
fixed.  Hence, the generalized Bekenstein bound can be violated.

As $M$ is increased above the range (\ref{eq-pencilrange}), states
with higher $(k,l)$ and multiparticle states become allowed.  But
Eq.~(\ref{eq-nmpencil}) still gives a lower bound on ${\cal N}(M)$.
It follows that an aspect ratio of
\begin{equation}
\epsilon_{\rm vio} = M \exp (-\pi^2 M)
\end{equation}
is sufficient to violate the bound for any $M\gg 1$.  This means that
one side of the box is exponentially longer than the other two sides.

For smaller values of $M$, the required aspect ratio does become
practically feasible, though $\epsilon$ still has to be quite small to
cause violation.  To illustrate this, let us not fix $M$ but find the
thickest system (the largest $\epsilon$) that allows violation for
{\em some} choice of $M$.  We take Eq.~(\ref{eq-nmpencil}) to be exact
and assume that ${\cal N}\gg 1$.  After the calculation we will
demonstrate that this approximation is self-consistent.

From Eq.~(\ref{eq-nmpencil}), we see that the bound is violated if
\begin{equation}
\epsilon^2 < (M^2 - 2) \exp(-2\pi^2 M).
\end{equation}
The right hand side is maximal for
\begin{equation}
M(\epsilon_{\rm vio}^{\rm max}) = \frac{1}{2\pi^2} +
\sqrt{2+\frac{1}{4\pi^4}} \approx 1.47 .
\end{equation}
Hence the largest aspect ratio leading to violation takes the value
\begin{equation}
\epsilon_{\rm vio}^{\rm max} \approx 2.01 \times 10^{-7}.
\end{equation}

The critical mass $M(\epsilon_{\rm vio}^{\rm max})$ falls well below
the value $\sqrt{5}$ at which Eq.~(\ref{eq-nmpencil}) becomes an
underestimate.  We also note that ${\cal N}(M)$ grows far less rapidly
than $\exp(\pi^2 M)$ for $M> M(\epsilon_{\rm max})$ and that the bound
is obeyed in the thermodynamic limit.  Finally, we may verify that
${\cal N}\approx 1/\epsilon \gg 1$.  Hence our above approximation is
self-consistent and the bound cannot be violated for any values of
$\epsilon$ exceeding $\epsilon_{\rm vio}^{\rm max}$.  We conclude that
pencil-shaped cavities whose length exceeds their diameter by a factor
at least of order $10^7$ admit violations of the bound at least for
low values of $M$ of order the inverse width of the cavity.

\subsubsection{Pancake-shaped cavity}
\label{sec-pancake}

Now consider a thin, pancake-shaped cavity with $a \ll b$ and $a\ll
c$, and define the small aspect ratios $\delta\equiv a/b$ and
$\epsilon\equiv a/c$.  Taking $a=\pi$, Eq.~(\ref{eq-omegas})
simplifies to
\begin{equation}
\omega = \left( k^2+l^2 \delta^2 +m^2 \epsilon^2 \right)^{1/2}.
\end{equation}
Again there is a gap, $M\approx 1$, above which densely spaced states
become allowed.  Below $M\approx 2$ all non-vacuum states are single
particle states with $k=1$.  The number of states below $M$ is given
by
\begin{equation}
{\cal N}(M) = \frac{\pi(M^2-1)}{4\delta\epsilon}
\label{eq-nmpancake}
\end{equation}
to an approximation which is good for $\delta\epsilon\ll M^2-1\leq 1$.

Again, we find that the entropy grows logarithmically with the
transverse area $A$:
\begin{equation}
S \approx -\log\delta\epsilon = \log \frac{A}{a^2}.
\label{eq-spancake}
\end{equation}
Since the generalized Bekenstein bound is independent of $A$, it can
be violated for sufficiently large $A$.  This remains true even at
larger values of the energy, $M>2$, when the right hand side of
Eq.~(\ref{eq-nmpancake}) becomes only a crude lower bound on ${\cal
N}$.

It is interesting to ask ``how thin'' the cavity must be so that the
bound can be violated for some $M$.  The bound is violated if
\begin{equation}
\delta\epsilon < \frac{\pi}{4} (M^2-1) \exp(-\pi^2 M).
\end{equation}
The right hand side is maximal for $M= \pi^{-2} + (1+\pi^{-4})^{1/2}
\approx 1.11$.  Hence the bound can be violated if
\begin{equation}
\delta \epsilon < 3.19 \times 10^{-6}.
\end{equation}
In particular, for a square cavity ($\delta=\epsilon$), violations
occur if the width-to-transverse-length ratio is at most
\begin{equation}
\epsilon_{\rm vio}^{\rm max} \approx 1.78 \times 10^{-3}.
\end{equation}

\subsubsection{Summary}

We have considered thin rectangular cavities whose sidelengths obey
$a\leq b\leq c$ and $bc\gg a^2$.  We expect our conclusions to apply
qualitatively for more general cavities with transverse area much
larger than the smallest dimension squared ($A\gg a^2$).

Then Eqs.~(\ref{eq-spencil}) and (\ref{eq-spancake}) suggest that the
entropy of any thin system at fixed low energy ($M\approx 1/a$) generally
grows with the transverse area as
\begin{equation}
S\approx\log\frac{A}{a^2},
\label{eq-loga}
\end{equation}
whereas the right hand side of Eq.~(\ref{eq-gbb}) is independent of
$A$ and of order unity.  For systems with dimensions of order
$a^2/A\approx 10^{-6}$ or thinner, the entropy in Eq.~(\ref{eq-loga})
therefore violates the bound.

\subsection{Comparing and combining problems}
\label{sec-mix}

Among the three problem types we present here, two (species and
Casimir) arose already in the context of the original Bekenstein bound
(\ref{eq-bb}), in which the circumferential diameter is used instead
of the (smallest) width $a$.  The different definition of the relevant
length scale means that the original bound is immune to the transverse
problem, which arises only for the generalized Bekenstein bound,
Eq.~(\ref{eq-gbb}).

The different problems can exacerbate each other when they are
combined.  For example, one can trade high species number for thinness
in concocting apparent counterexamples to the bound.  For $Q>1$,
violations occur already at larger (``less thin'') values of
$\epsilon$ than those found in Sec.~\ref{sec-transverse}.

Similarly, the Casimir problem can exacerbate the transverse problem.
Suppose that $E_C<0$ for a given combination of fields in the thin
limit ($A\to\infty$ as $a$ is held fixed).  Take $Q=1$ and recall the
pencil-shaped cavity studied in Sec.~\ref{sec-transverse} and the
conventions used there.  The vacuum energy receives a Casimir shift of
order $\eta/\epsilon$.  For the pancake shape, the shift is of order
$\eta/\delta\epsilon$.  In either case the shift can be made
arbitrarily large, and the energy of many excited states will be
negative, or positive but too small for the bound to be valid.

The species problem, like the thin-cavity problem, may also be
exacerbated by Casimir energy.  If the Casimir energy is negative for
some field, then it can be made arbitrarily large in magnitude by
including $Q$ such fields.  In such cases, ${\cal N}(M)$ grows with
$Q$ not only due to the greater degeneracy of energy levels, but also
due to the negative contribution from $E_{\rm C}$.

\section{Confinement energy}
\label{sec-wall}

So far, our analysis of field systems has assumed Dirichlet boundary
conditions without including the matter required to enforce them.  But
a free field cannot be legislated to vanish outside a given region of
space; physically, it must be forced to do so by couplings to other
matter or to itself.  If the Bekenstein bound applies only to complete
systems, then our analysis so far will have fallen short.

For example, the electromagnetic field modes in a cavity will vanish
on the boundary only if a sufficient number of electric charge
carriers are available in the boundary material.  Their density and
coupling strength will also determine the depth of penetration of
field modes into the surrounding conductor, and thus its minimum
width.  Bekenstein~\cite{Bek00b} recently showed that the extra energy
contributed by such charge carriers suffices to restore the validity
of the bound (\ref{eq-bb}) in apparent counterexamples considered by
Page~\cite{Pag00a,Pag00b}.\footnote{One of Page's
counterexamples~\cite{Pag00b} to Eq.~(\ref{eq-bb}) is really a more
complicated relative of the pencil problem of Sec.~\ref{sec-pencil}.
Our setup is simpler because it is designed to challenge only the
stronger version of Bekenstein's bound, Eq.~(\ref{eq-gbb}).}

The purpose of this section is to generalize this type of argument.  A
specific model is constructed in order to estimate a lower bound
on the energy cost of confining fields to a region of width $a$.  We
then revisit our earlier examples and ask whether the extra energy
resolves the apparent counterexamples found in Sec.~\ref{sec-bad}.

\subsection{Energy of a confining background field}
\label{sec-background}

\subsubsection{A background field model of confinement}

Let us begin by estimating what it takes to confine a single scalar
field $\phi$ to a cavity (we will soon generalize this to multiple
species).  We model the confining matter by a turning on a static
background field $\sigma(x,y,z)$ with standard kinetic term
$(\nabla\sigma)^2$, which couples to the field $\phi$ through a term
$\lambda\phi^2\sigma^2$.  We must assume that the dimensionless
coupling obeys
\begin{equation}
\lambda<1,
\label{eq-lambda}
\end{equation}
since strong coupling leads to large corrections; in particular,
one-loop contributions to the energy would not be under control.  We
will allow $\sigma$ to be any function of the spatial coordinates, but
we insist on adding the integrated gradient energy density of the
$\sigma$ field to the total energy:
\begin{equation}
E = E_\phi + E_{\rm C} + E_\sigma,
\end{equation}
where
\begin{equation}
E_\sigma =  {1\over 2}\int d^3x (\nabla_i \sigma) (\nabla^i \sigma).
\end{equation}

Implicit in this model is a distinction between an active field and a
confining, non-dynamical background field.  This is obviously somewhat
artificial.\footnote{Depending on the Lagrangian, the model can be
physically realized by solitons.  Nonperturbative solutions generate a
potential for perturbative states, some of which may be bound.
Application of the Bekenstein bound in this context was investigated
in Ref.~\cite{BekSch90}.  We emphasize, however, that our eventual
definition of $S$ in Eq.~(\ref{eq-di}) refers to any discrete
eigenstate of the spectrum, i.e., not only to bound states arising in
a background potential but also to those arising simply by
perturbative interactions.---I wish to thank J.~Bekenstein for
pointing out Ref.~\cite{BekSch90} to me.}  After all, dynamical fields
are quite capable of confining themselves in interacting field
theories, and in Sec.~\ref{sec-conclusions}, a revised definition of
entropy will be given which refers to complete systems from the start.
For now, our task is merely to estimate how much energy it takes to
patch up examples that are physically incomplete field systems.

We begin with a set of field modes obeying sharp boundary conditions,
as in Secs.~\ref{sec-ok} and~\ref{sec-bad}.  Next, we modify the
system by couplings to a background field, enough to justify the
spatial confinement of modes, but without significantly changing the
number of states.  Then we estimate the minimum energy contributed by
the confining matter.  Finally, we show that due to this extra energy,
our naive conclusion that Bekenstein's bound is violated was not
self-consistent.

The Fock space construction begins as usual by expanding $\phi$ into
energy eigenmodes,
\begin{equation}
\phi(x,y,z,t) = \sum_{\mathbf k} a_{\mathbf k} 
u_{\mathbf k}(x,y,z) e^{i \omega t} + \mbox{c.c.},
\end{equation}
which obey the Schr\"odinger equation
\begin{equation}
(\nabla_i \nabla^i + \lambda \sigma^2 ) u_{\mathbf k} =
 \omega^2 u_{\mathbf k}.
\end{equation}  

Acting on the vacuum, the operator $a_{\mathbf k}^\dagger$ creates a
one particle state of energy $\omega({\mathbf k})$.  We call this
particle {\em confined\/} to a region ${\cal V}$ if the support of the
corresponding mode function is concentrated in that region, i.e., if
\begin{equation}
\int_{\cal V} d^3x\, u_{\mathbf k} u^*_{\mathbf k} 
\approx \int_{\bf R^3} d^3x\, u_{\mathbf k} u^*_{\mathbf k}.
\label{eq-supp}
\end{equation}
More general energy eigenstates of the Fock space are constructed by
acting with a variety of creation operators $a_{{\mathbf k}_1}$,
$a_{{\mathbf k}_2}$, etc., one or multiple times.  We call such a
state {\em confined\/} to ${\cal V}$ if and only if all of its
particles are confined to ${\cal V}$.

\subsubsection{Estimating the confinement energy}

To achieve sharp boundary conditions for the field $\phi$, the
background field $\sigma$ would have to form an infinite potential
well: inside ${\cal V}$, $\sigma=0$; outside, $\sigma\to\infty$.  This
requires taking a limit in which $\sigma$ changes by an infinite
amount over zero spatial distance, so that $E_\sigma$ will diverge.
Hence, the cavities studied in Secs.~\ref{sec-ok} and
Sec.~\ref{sec-bad} would actually have infinite total energy.

In a cavity, all but a finite number of modes have $\omega>M$ and
hence would not count towards the entropy $S$ in any case.  There is
no point in confining such modes.  Moreover, there is no need to
suppress the amplitude of modes sharply; particles are considered
well-localized to a region $\cal V$ even if there is a tiny
probability to find them elsewhere.

Hence, consider instead a potential of finite height $\sigma_0>0$,
such as a finite well with $\sigma=0$ inside some compact region
${\cal V}$ and $\sigma=\sigma_0$ otherwise.\footnote{We are choosing
the region entering the Bekenstein bound to coincide with the
potential well.  This is not necessary, but it is the most interesting
choice for the purposes of challenging the bound.}  This potential can
only bind $\phi$-particles whose momenta have magnitude less than
\begin{equation}
k_0=\sigma_0\sqrt{\lambda}
\label{eq-klambda}
\end{equation}
in the rest frame of the background potential.  Modes with larger
momentum will not be bound.  For rectangular ${\cal V}$, the bound
modes will be roughly of the form of Eq.~(\ref{eq-rmds}), except that
their amplitude will not turn off sharply at the potential step.
Instead, the amplitude develops an exponential tail of width $(k_0^2 -
k^2)^{-1/2}$.

A finite potential well still has infinite gradient energy.  However,
the modes change negligibly if the discontinuous jump of $\sigma$ is
slightly smeared.  In order to reduce the gradient energy $E_\sigma$
as much as possible, let us smear $\sigma$ over the entire region
${\cal V}$, i.e., let $\sigma$ rise continuously from $0$ in the
center of ${\cal V}$ to $\sigma_0$ at the boundary of ${\cal V}$;
outside ${\cal V}$, we still take $\sigma=\sigma_0$, which costs no
gradient energy at all.

We can now estimate a lower bound on the energy required to confine a
single field species to a region of width $a$ and transverse
cross-section $A$.  The gradient energy density is at least of order
$\sigma_0^2/a^2$, since $\sigma$ has to drop to $0$ and rise back to
$\sigma_0$ over a distance $a$.  Integration over the volume $Aa$
yields
\begin{equation}
E_\sigma \gtrsim \frac{A\sigma_0^2}{a}\gtrsim \frac{A k_0^2}{a},
\label{eq-wall}
\end{equation}
where Eqs.~(\ref{eq-lambda}) and (\ref{eq-klambda}) have been used.

\subsubsection{Estimating the entropy}

We now propose to test the generalized Bekenstein bound as follows.
Pick a background field $\sigma (x,y,z)$ with finite gradient energy
$E_\sigma$.  For any $M\geq E_\sigma$ and any spatial region ${\cal
V}$, let ${\cal N}(\sigma,{\cal V};M)$ be the number of energy
eigenstates in the Fock space which are confined to ${\cal V}$ and
whose total energy $E$ does not exceed $M$.  Let $a$ be the width of
${\cal V}$ as defined in the introduction (e.g., if ${\cal V}$ is a
rectangular box, $a$ can be chosen to be its smallest dimension).  The
entropy is defined microcanonically as
\begin{equation}
S(M) \equiv\log {\cal N}(\sigma,{\cal V};M).
\end{equation}
The bound asserts that
\begin{equation}
S(M) \leq \pi M a
\end{equation}
for all $\sigma$, ${\cal V}$, and $M$.

To calculate the entropy, one would need to find the eigenmodes of the
smeared potential and count up all the confined Fock space states with
total energy below $M-E_\sigma$.  However, it is usually unnecessary
to work out explicit mode solutions; a reasonable upper bound for the
entropy can be estimated from the known solutions for an infinite well
occupying the same region ${\cal V}$.  Infinite well modes with
momenta below $k_0$ will remain confined in a finite well with
$\sigma=\sigma_0$ outside ${\cal V}$.\footnote{Bound states which are
almost scattering states must also be excluded because they have large
exponential tails outside ${\cal V}$ and so fail to satisfy
Eq.~(\ref{eq-supp}) (unless the characteristic length entering the
bound is increased accordingly).  E.g., one-dimensional modes with
momentum $k$ have tails of size $(k_0^2 - k^2)^{-1/2}$, which must not
be much larger than the width of the potential valley; hence $k$ must
not be allowed to approach $k_0$ to closely.}  Under further
deformation to a smooth potential valley, one would expect at most
order one changes in the energy and size of most bound state modes.
Also, since both quantities should increase under this deformation, we
will be overestimating the entropy or underestimating the right hand
side of the bound, which is permitted if we nevertheless find the
inequality to be satisfied.

\subsection{Examples}

We will now revisit the various examples studied earlier, in order to
include confinement in our analysis.  We will find that confinement
energy acts complementary to the field energy.  In examples where the
bound was already found to be easily satisfied, corrections from
boundary effects are small, validating the earlier analysis.  For
examples where saturation of the bound was approached, the boundary
effects lead to corrections of order one.  Most importantly, for the
problematic examples that appeared to violate the bound, we will find
that confinement is actually the dominant source of energy, rendering
our earlier analysis invalid.

\subsubsection{High temperature limit}
\label{sec-rcube}

As our first example, let us revisit the high temperature limit of a
scalar field in a cavity, studied in Sec.~\ref{sec-thermal}.  This is
the regime
\begin{equation} 
Ta\gg 1,
\end{equation}
in which thermodynamic approximations are valid.  We set the number of
species to one ($Q=1$), postponing the analysis of $Q>1$ to
Sec.~\ref{sec-rspecies}.

In order to apply the general result for confinement energy,
Eq.~(\ref{eq-wall}), we need to know how large $k_0$ has to be.  In
the thermal regime, the typical $\phi$-modes have energy and momenta
of order $T$, so confinement is achieved by choosing $k_0=T$.  Since
$a$ is the shortest width, we have $A\geq a^2$, and
Eq.~(\ref{eq-wall}) implies
\begin{equation}
E_\sigma\gtrsim T^2 a.
\label{eq-esigmacube}
\end{equation}

As a function of $T$, the entropy is
\begin{equation}
S(T) \approx (Ta)^3
\end{equation}
whereas the right hand side of the generalized Bekenstein bound
contains two terms
\begin{equation}
Ea \approx E_\phi a + E_\sigma a \approx (Ta)^4 + (Ta)^2
\end{equation}
by Eqs.~(\ref{eq-et}) and (\ref{eq-esigmacube}).

At leading order in the high temperature limit, we see that the ratio
\begin{equation}
{S\over Ea} \approx {1\over Ta}
\end{equation}
is not affected by confinement.  The Bekenstein bound is satisfied
with the same comfortable factor $(Ta)$ to spare that we already
determined in Sec.~\ref{sec-thermal}.  This is a sensible result.  If
a system is much larger than the typical wavelengths it contains, then
the contribution of boundaries to the energy should not have to be
large.

\subsubsection{Saturation limit}

Next, let us turn to the limit in which saturation of the bound may be
approached, as exemplified in Sec.~\ref{sec-cube}.  This occurs for
systems whose various dimensions are all roughly equal to a single
length scale $a$.  Examples include a cubic (or spherical) cavity of
side length (or diameter) $a$ confining a single massless scalar.
Saturation is approached as the maximum total energy, $M$, is lowered
to become of order $1/a$.  Then the cavity is occupied by a small
number of quanta whose typical wavelength is comparable to the cavity
diameter, $a$.  Since the entropy is of order one, the canonical
ensemble cannot be used as an approximate substitute for the
microcanonical counting of states we employ.

Recall that the total energy of the $\phi$-states, in this limit, is
comparable to the energy of an individual $\phi$-particle, i.e.,
\begin{equation}
E_\phi \approx 1/a.
\end{equation}
Hence $E_\phi a$ is of order unity, like the entropy.  The momenta of
confined $\phi$ particles are at least $1/a$ by the uncertainty
relation. Thus, Eq.~(\ref{eq-wall}), with $A\approx a^2$, implies
\begin{equation}
E_\sigma \gtrsim 1/a.
\end{equation}
Hence, the confinement contribution to the right hand side of the
bound is of the same order as that of the active field $\phi$.
Compared to the analysis without attention to confinement, the bound
is now satisfied with an extra factor $\gtrsim 1$ to spare.

We conclude that upon inclusion of confinement energy, systems with
very few quanta can still come reasonably close to saturating the
bound.  In this limit, however, the boundary contribution begins to
dominate over the field energy.  This would appear to make it even
more difficult to devise simple systems in which the bound is
saturated exactly.

\subsubsection{Resolving the Casimir problem}
\label{sec-rcasimir}

If the vacuum state had negative energy $E_{\rm C}$, the Bekenstein
bound would be automatically violated for negative values of $M\geq
E_{\rm C}$ (Sec.~\ref{sec-casimir}).  The calculation of Casimir
energies is often carried out with sharp boundary conditions
separating bound states inside the region ${\cal V}$ from the
continuous spectrum outside.  As we have discussed above, perfect
confinement gives rise to infinite barrier energy.  However, imperfect
barriers also lead to Casimir energies, and Eq.~(\ref{eq-ecest})
retains its general form even with an ultraviolet cutoff.  Hence, we
will only assume that the background potential confines at least one
particle to a region of width $a$.

As in the previous example, the uncertainty principle demands that the
barrier height, $k_0$, must be greater than $1/a$.  By
Eq.~(\ref{eq-wall}), the confining matter therefore has at least
energy
\begin{equation}
E_\sigma \gtrsim 1/a.
\end{equation}
Comparison with Eq.~(\ref{eq-ecest}) shows that the Casimir energy is
parametrically smaller in magnitude than the positive confinement
energy~\cite{Bek82}.  This suggests that the total energy is always
positive---a result which one expects for complete systems on physical
grounds in any case~\cite{Bek94}.

\subsubsection{Resolving the species problem}
\label{sec-rspecies}

At fixed energy, the entropy of weakly interacting fields confined to
a fixed compact region will increase monotonically with the number of
field species.  This constitutes the species problem
(Sec.~\ref{sec-species}).  We will assume a region of regular shape,
$A\approx a^2$, for simplicity.

If a type of particle is not confined to the given region, its modes
do not contribute to the entropy.  For example, one might build a box
that confines photons and electrons, but not neutrinos; in that case,
one would not include neutrino states in $S$.  Hence, we must adapt
our estimate of confinement energy, Eq.~(\ref{eq-wall}), to the
problem of confining multiple fields.  We will focus on two simple
methods which we expect to be representative, and which give the same
answer.

One possibility is to couple each active field $\phi_q$ to a different
background potential $\sigma_q$ via coupling terms
\begin{equation}
\sum_q \lambda_q \phi_q^2 \sigma_q^2.
\end{equation}
Since we must require that none of the couplings are strong
($\lambda_q<1$ for all $q$), each background field $\sigma_q$ will
contribute gradient energy
\begin{equation}
E_{\sigma_q} \approx \int(\nabla\sigma_q)^2\gtrsim 1/a.
\end{equation}
Hence the total confinement energy obeys
\begin{equation}
E_{\sigma}\gtrsim Q/a.
\label{eq-eqw}
\end{equation}

A second possibility is to couple several or all fields $\phi_q$ to
the same background potential $\sigma$ via terms $\lambda \sum_q
\phi_q^2 \sigma^2$.  Then a potential $\sigma$ that confines certain
modes of $\phi_1$ to a compact region will confine the corresponding
modes of all other fields.  Hence, it would appear that the
confinement energy will be independent of the number of fields, $Q$.
However, radiative corrections to the $\sigma$ propagator will now be
larger, $Q\lambda$, because $Q$ fields are running in the loop.
Hence, Eq.~(\ref{eq-lambda}) is no longer sufficient to bring such
corrections under control.  Instead, we must demand that
\begin{equation}
\lambda<1/Q.
\end{equation}
This modifies our estimate for the confinement energy:
Eq.~(\ref{eq-wall}) must be replaced by
\begin{equation}
E_\sigma \gtrsim \frac{A\sigma_0^2}{a}\gtrsim \frac{Q A k_0^2}{a}.
\label{eq-wall2}
\end{equation}
Compared to the previous paragraph, we hoped to save gradient energy
by coupling all active fields to the same background field.  But at
the same time we gained gradient energy because the couplings are now
restricted to be weaker, requiring a greater potential height
$\sigma_0$.  The result is the same in both cases: Eq.~(\ref{eq-eqw}).

In order to obtain apparent violations in Sec.~\ref{sec-species}, the
number of species had to be chosen very large: $Q\gg 1$.  Since
$E_\phi\approx 1/a$ in the mass range where the bound is most easily
threatened, the total energy will be dominated by confinement effects
for $Q\gg 1$, so that the right hand side of Bekenstein's bound is at
least of order $Q$.  This conservative estimate increases linearly
with $Q$, while the entropy on the left hand side grows only
logarithmically.  Hence, the bound, far from being violated, actually
becomes more comfortably satisfied when the number of species is
increased in the background field model.

Physically, this result is easily understood as follows.  Start with a
well-localized system (say, a few atoms forming a small crystal) and
increase the species number at fixed energy.  If the new species all
couple with the same strength as the original ones, then the system
will develop strong 't~Hooft couplings.  It will be transformed by
radiative corrections, rendering its size and entropy difficult to
calculate and invalidating the naive argument for a violation of the
bound.  If, on the other hand, some of the new species are more weakly
coupled than the original ones, so as to avoid strong 't~Hooft
coupling, then the system will delocalize.  The more species, the
weaker the required couplings, so that the system grows in size, and
the bound increases more rapidly with species number than the entropy
does.

\subsubsection{Resolving the transverse problem}
\label{sec-rtransverse}

Neglecting confinement, we noted in Sec.~\ref{sec-transverse} that the
generalized (though not the original) Bekenstein bound is violated by
``thin'' systems, whose width $a$ is much smaller than its transverse
size.  Choosing the maximum energy $M$ to be of order $1/a$, one finds
that only one-particle states are allowed, whose number grows as a
power law with the transverse area.  Hence, the entropy grows
logarithmically with the transverse area; see Eq.~(\ref{eq-loga}).
Yet the right hand side of the bound is fixed and of order unity.

The estimate of the confinement energy is simple.  We need only
consider the contributions from the transverse surfaces (e.g., the
capacitor plates) to Eq.~(\ref{eq-wall}).  This contribution dominates
over the other boundaries of the system for several reasons: (i) the
transverse boundaries have the largest area, $A$; (ii) they set the
required barrier height $\sigma_0$, because $1/a$ is the largest
momentum component; and (iii) they contribute the largest gradient
energy density to $(\nabla\sigma)^2$ because $a$ is the shortest
dimension.  Substituting $k_0\approx 1/a$ in Eq.~(\ref{eq-wall}), we
find for the total confinement energy:
\begin{equation}
E_{\sigma} \gtrsim \frac{A}{a^3}
\end{equation}

The confinement contribution to the right hand side of the entropy
bound is $E_\sigma a$.  By the above inequality, this is at least
$A/a^2$ and scales linearly with the transverse area.  In a regular
system, for which $A\approx a^2$, we have already verified above that
the bound is safe.  As we increase the transverse size $A$ at fixed
width $a$, the right hand size of the bound thus grows more rapidly
than the entropy $S$ on the left hand side.  Hence the bound remains
valid for thin systems.  With confinement effects included, the ratio
$S/(\pi Ma)$ actually {\em decreases\/} as the transverse size grows:
the increase in entropy is more than compensated by larger energy
needed to enforce boundary conditions along a bigger surface.

\section{Containment energy}
\label{sec-pressure}

In the previous section we argued that confinement---the need to
enforce boundary conditions on localized energy eigenmodes---adds
energy to systems which naively appear to violate Bekenstein's bound,
ensuring that the bound is in fact satisfied.  In this section we
explore an alternative source of extra mass that has been
suggested~\cite{Bek82} as a contribution protecting the Bekenstein
bound.  We will ultimately dismiss this contribution as both
insufficient and inessential.

The energy of any quantum state in a cavity generally varies with the
volume of the cavity.  Indeed, the presence of Casimir energy ensures
that this is true even for the vacuum state.  This implies that the
system exerts pressure on its boundary.  The model of the previous
section focussed on the need for interactions that keep field modes
confined to a given region.  It did not explicitly address the
requirement that the cavity must be mechanically stable against
implosion or expansion, which is clearly necessary if energy
eigenstates (which are static) are to be considered.  We will now
devise a different model for estimating additional energy, which
focusses not on interactions but instead on the mechanical properties
of a container wall counteracting the pressure.

As in Sec.~\ref{sec-ok} and~\ref{sec-bad} (but in contrast with
Sec.~\ref{sec-wall}), we consider free fields obeying sharp boundary
conditions.  However, we will now take into account the pressure or
suction on the boundary surface.  We surround the system by a thin
wall.  This container must be sufficiently rigid to withstand the
range of pressures caused by the states one wishes to include in the
entropy $S$.  We estimate a lower bound on the mass of the required
container.  This mass can be expressed as a function of the pressure
and the shape of the boundary, generalizing earlier estimates by
Bekenstein~\cite{Bek82}.  Then we investigate how this additional
energy affects the validity of Bekenstein's bound in various examples.

\subsection{Mass of a thin container}

Consider a region ${\cal V}$ surrounded by a stationary wall, a
container which stabilizes the matter occupying ${\cal V}$ against
pressure or suction.  Let us approximate the container as a
codimension-one source of stress-energy, with distributional stress
tensor
\begin{equation}
T^{ab} = S^{ab} \delta[r-r_0(\vartheta,\varphi)],
\end{equation}
where $r$ is a Gaussian normal coordinate along geodesics generated by
the vector field $n_a$ orthogonal to the wall; $\vartheta$ and
$\varphi$ are the remaining two spatial coordinates.

The extrinsic curvature of the wall in the spacetime is given by the
second fundamental form,
\begin{equation}
K_{ab} =\nabla_a n_b.
\end{equation}
The only nonvanishing entries of $K_{ab}$ are the spatial components
tangential to the container, which we indicate by indices $i\ldots$
running over $\vartheta$ and $\varphi$.  Let $P$ be an arbitrary point
on the wall.  By suitable rotation of the coordinate system about
$n_a$, $K_{ij}$ can be diagonalized:
\begin{equation}
K_{ij} = \mbox{diag}(K_{\vartheta\vartheta}, K_{\varphi\varphi}).
\end{equation}
The diagonal components are the inverse curvature radii of the wall in
the principal directions at $P$; negative entries correspond to
concave directions.

Let $p$ be the pressure exerted by the enclosed bulk system on the
container at the point $P$.  This pressure results in a force $p\,
d\vartheta\, d\varphi$ normal to a surface element $d\vartheta\,
d\varphi$.  The wall exerts a force $S^{ij} K_{ij}\, d\vartheta\,
d\varphi$ orthogonal to itself.  Thus, the forces cancel if
\begin{equation}
p = -S^{ij} K_{ij}.
\label{eq-pressure}
\end{equation}

Here we are given a problem where $p$ and $K_{ij}$ are specified, and
the task is to find the most energy-saving material by which the
system can be enclosed and stabilized.  That is, we must find a tensor
$S^{ij}$ that satisfies Eq.~(\ref{eq-pressure}) while minimizing the
surface energy density, $S^{00}$.  The latter is constrained by the
dominant energy condition~\cite{Wald}, which implies for each (fixed)
spatial index $i$ that
\begin{equation}
S^{00} \geq 
\sqrt{(S^{i i})^2 + \sum_{j\neq i} (S^{ij})^2 + (S^{0j})^2}.
\label{eq-decij}
\end{equation}

Recall that the container is taken to be a thin wall, i.e., $S^{ab}
n_{a} =0$.  In the frame where $K_{ij}$ is diagonal, the off-diagonal
components of $S^{ab}$ do not contribute to the pressure balance in
Eq.~(\ref{eq-pressure}).  Moreover, by Eq.~(\ref{eq-decij}), turning
on any off-diagonal components can only increase the minimum energy
density of the container.  Hence $S^{00}$ is minimized by choosing a
stress-free, isotropic\footnote{More precisely, we take all of the
diagonal components to be mutually equal in magnitude.  The sign of
each $S^{i i}$ must be chosen opposite to the sign of the
corresponding $K_{i i}$.  If the signs vary, the material will not be
isotropic.} wall material:
\begin{equation}
|S^{ij}|= \frac{|p|}{\hat{K}}\, \delta^{ij},
\end{equation}
where
\begin{equation}
\hat K=\mbox{tr}\, |K_{ij}|.
\end{equation}

Then we find for the energy per surface area of the container:
\begin{equation}
S^{00} \geq  {|p|\over \hat{K}}.
\label{eq-s00}
\end{equation}
Physically, this means that the surface density is at least of order
the pressure times the smallest curvature radius at each point.  This
lower bound is both necessary and sufficient for the dominant energy
condition, so it cannot be improved.  The total container mass
therefore obeys
\begin{equation}
E_{\rm cont} = \int d^2x \sqrt{h} S^{00} \geq \int d^2x \sqrt{h}
{|p|\over \hat{K}}.
\label{eq-ep}
\end{equation}

\subsection{Examples}

Having established a lower bound on the container mass, we may
consider the effect of this additional term on the validity of the
generalized Bekenstein bound.  Thus, let us refine our definition of
energy to include the pressure term,
\begin{equation}
E \equiv E_\phi + E_{\rm C} + E_{\rm cont}.
\label{eq-nmdef3}
\end{equation}
We will begin by reconsidering examples in which the bound was earlier
found to be satisfied (Sec.~\ref{sec-ok}); then we will return to the
problematic examples of Sec.~\ref{sec-bad}.

\subsubsection{High-temperature limit and saturation limit}

In Sec.~\ref{sec-thermal} we considered $Q$ scalar fields in a cavity
at temperature $T\gg 1/a$.  For simplicity, let us now assume that the
cavity is evenly shaped, say spherical with radius $R$.  We work up to
factors of order unity.  The bulk
pressure is
\begin{equation}
p\approx QT^4,
\end{equation}
and the trace of the extrinsic curvature is
\begin{equation}
\hat K \approx 1/R.
\end{equation}
Thus, the mass of the containing wall obeys
\begin{equation}
E_{\rm cont}\gtrsim Q R^3 T^4
\end{equation}
by Eq.~(\ref{eq-ep}).  

Comparison with Eq.~(\ref{eq-et}) shows that the wall mass is at least
of the same order as the field energy it contains.  Unlike the
confinement energy, the containment contribution does not become
negligible in the high-temperature limit.  At all temperatures, its
inclusion significantly increases the margin by which the bound is
satisfied.

This is true also in the saturation limit, which can be approximately
treated within the above argument, by reducing the temperature to
$T\approx 1/R$.  Instead of approaching saturation to within a factor
of 10 or 100, the gap becomes correspondingly larger upon inclusion of
$E_{\rm cont}$.

\subsubsection{Casimir problem}

If it turned out that containment energy necessarily renders the total
energy of the system positive, i.e., if
\begin{equation}
E_{\rm cont} \geq |E_{\rm C}|,
\end{equation}
then one could argue that the Casimir problem is resolved.  In fact,
however, there is a simple example in which this cannot be shown.

Consider a spherical cavity containing some field(s) with negative
Casimir energy (for example, a massless scalar obeying Neumann
boundary conditions~\cite{BorMoh01}).  In the vacuum, the bulk
pressure is
\begin{equation}
|p|=\frac{|E_{\rm C}|}{4\pi R^3}.
\end{equation}
The extrinsic curvature is
\begin{equation}
\hat K = 2/R,
\end{equation}
so that the surface density of the containing wall obeys
\begin{equation}
S^{00}\geq \frac{|E_{\rm C}|}{8\pi R^2}.
\end{equation}
Hence, the wall mass satisfies merely
\begin{equation}
E_{\rm cont} \geq |E_{\rm C}|/2.
\label{eq-contec}
\end{equation}
For fields with $E_{\rm C}<0$, we are thus unable to conclude that
$E_{\rm C}+ E_{\rm cont}\geq 0$.

Of course, Eq.~(\ref{eq-contec}) is not generic; it depends on our
specific choice of a spherical container, which minimizes wall energy.
For many other shapes, containment energy does compensate the Casimir
energy.  (For example, for a long cylinder one finds that $E_{\rm
cont} \geq 2 |E_{\rm C}|$.)  To realize $E_{\rm cont}+|E_{\rm C}|<0$,
one would require extremely stiff layers ($S^{ii}>S^{00}/2>0$),
which---though allowed by the dominant energy condition---may not
actually occur in nature.  Nor should Eq.~(\ref{eq-contec}) be
interpreted as a claim that a complete stationary system could have
negative total energy due to a Casimir contribution.

The point here is merely that containment is not enough to demonstrate
positive total energy; hence, it does not by itself resolve the
Casimir problem of the Bekenstein bound.

\subsubsection{Species problem}
\label{sec-nspecies}

Bekenstein~\cite{Bek82,Bek83} has argued that containment resolves the
species problem as follows.  Each species confined to the region
${\cal V}$ contributes a Casimir energy $ E_{{\rm C},q}$, whose
pressure or suction must be compensated by a container wall.  The wall
energy will thus depend roughly linearly on the number of species,
whereas the entropy grows only logarithmically.  At the large species
numbers necessary for apparent violations of the bound, the wall
energy will be overwhelming.

This argument is not fully satisfactory.  Since the Casimir energy has
no definite sign, one could chose a combination of fields whose
Casimir energies mostly cancel, yielding only a small correction to
the energy of the first single partice state.  Then the species
problem would arise as in Sec.~\ref{sec-casimir}.  Though containment
energy does resolve the species problem in generic examples, this
potential loophole remains.

\subsubsection{Transverse problem}
\label{sec-ntransverse}

Returning to the rectangular cavities studied earlier, we note that
their boundary surfaces are flat, so that $S^{00} \to \infty$ by
Eq.~(\ref{eq-s00}).  However, the wall energy can be rendered finite
by allowing the boundary to bulge slightly, as long as this does not
significantly alter the shape of the cavity and invalidate the
analysis of field modes.  In particular, the width $a$ of the system
should still be approximately $a$.

Consider, for example, the two large, approximately flat, rectangular
surfaces of area $b c$ containing the pancake-shaped system of
Sec.~\ref{sec-pancake}.  The pressure or suction on this surface is at
least $|p|=|\eta|/a^4$ (from the Casimir pressure of the vacuum
state).  We may assume that $b\equiv a/\delta \leq c \equiv
a/\epsilon$.  A simple geometric argument shows that the principal
curvature radii of this surface must vastly exceed $a/\delta^2$ and
$a/\epsilon^2$, respectively, if the surface is to bulge by an amount
much less than $a$.  Thus we find that $\hat{K}\ll\delta^2/a$, and by
Eq.~(\ref{eq-s00}), the surface energy density must be much larger
than $|\eta|/(\delta^2 a^3)$.  Hence the total mass of the container
must be much larger than $|\eta|/(\delta^3\epsilon a)$.

Note that this contribution to the total energy scales like a power of
the transverse area, and dominates over $E_{\rm C}$ and $E_\phi$ for
thin systems ($\delta\ll 1$), in the regime where the transverse
problem arose ($E_\phi\approx 1/a$).  The corresponding growth of the
bound with transverse area outpaces the logarithmic divergence of the
entropy, removing the problem noted in Sec.~\ref{sec-pancake}.

As an example of a pencil-shaped system, consider a cylindrical cavity
of length $l\gg R$.  It has total curvature $\hat K = 1/R$ on its long
sides.  Because of their much smaller area, we neglect the two disks
that complete the boundary.  In the ground state the cylinder wall
sustains pressure $|p|= |E_{\rm C}| /\pi R^2 l$.  Hence the wall must
have a density $S^{00}\geq |E_{\rm C}|/\pi Rl$, and its total mass
obeys
\begin{equation}
E_{\rm cont} \geq 2 |E_{\rm C}|.
\end{equation}
This implies that $E_{\rm cont}+E_{\rm C}$, the total energy in the
ground state, is necessarily positive.  Though the energy grows only
linearly with transverse area in this case, the container mass
bolsters the bound sufficiently to stay ahead of the logarithmic
growth of the entropy.

However, a basic problem of the containment argument is that it relies
on the assumption that pressure on the container builds up at least in
proportion to the entropy.  As pointed out in Sec.~\ref{sec-nspecies},
this can be circumvented by cancelling off Casimir energies of
different signs.  More general problems arise if we do not insist that
the walls must be able to contain {\em all\/} states up to energy $M$,
but count only those states toward $S$ whose pressure is compatible
with the wall design.

For example, the Casimir force on a pair of conducting plates made
arbitrarily large by increasing the transverse area.  This force can
then be balanced by any one of a large number of excited states (since
only the wavelength along the short direction is fixed by the
cancellation requirement, leaving the transverse wavelength variable).
Thus it would seem that a container with little mass could hold a
large entropy.

Of course, this is false---but not because of pressure effects.  The
most general reason why one cannot construct massless capacitor plates
is that the necessary charge carriers cost energy.  But this is
precisely the confinement energy estimated in Sec.~\ref{sec-wall}.

\section{Bound states as complete systems}
\label{sec-conclusions}

In this section, we summarize the lessons learned from our analysis of
examples.  We conclude that interactions are crucial for the operation
of the Bekenstein bound: without them, the bound is trivial; but with
interactions properly included from the start, complete systems simply
correspond to bound states, whose number is measured by the entropy
$S$.

The analysis of Sec.~\ref{sec-pressure} has shown that realistic
containers add enough mass to protect Bekenstein's bound against many
problematic examples of the type constructed in Sec.~\ref{sec-bad}.
However, we have found that this is not guaranteed in general, and we
have pointed out some loopholes.  Confinement energy, as estimated in
Sec.~\ref{sec-wall}, appears to provide a more reliable mechanism
upholding the bound.

Demanding containment is not only insufficient---at present it is also
poorly motivated.  Containers did seem necessary in the context in
which Bekenstein's bound first arose: systems which are slowly lowered
towards a black hole, hovering just above the horizon before being
dropped in, are necessarily subjected to accelerations.  It is
therefore difficult to analyze the Geroch process independently of the
specific composition of the system, unless it is placed in a rigid
container for the duration of the gedankenexperiment.  

However, in the light of recent
developments~\cite{Tho93,Sus95,ceb1,ceb2,FMW,Bou03}, we take the point
of view that the Bekenstein bound, like the GSL, should be properly
regarded as a consequence of the covariant bound.  Then conditions
arising in the Geroch lowering process are not necessarily general
conditions for the validity of the Bekenstein bound.  In the recent
derivation of the Bekenstein bound from the GCEB~\cite{Bou03}, the
system follows a geodesic in flat space.  Hence, no separate container
is required, as long as the system holds itself together through
interactions.

We conclude that confinement energy, arising from the need for
interactions, is both the more plausible and the more effective
correction to consider when analysing apparent violations of
Bekenstein's bound: Firstly, it is well-motivated, since the energy
eigenstates of free fields cannot be localized to a finite region.
Secondly, it is successful, in the sense that the background field
model studied in Sec.~\ref{sec-wall} suggests that confinement
contributions invalidate the apparent counterexamples we presented in
Sec.~\ref{sec-bad}, without rendering the bound trivial.

Unfortunately, the background field model turns the application of
Bekenstein's bound into a two-step process: First, one works out the
entropy of an active free field with fixed boundary conditions; then,
one estimates the energy of an additional passive field needed for
enforcing those boundary conditions.  This is ugly, and moreover, both
steps are unphysical: the first, because truly free fields cannot obey
boundary conditions; the second, because there is no such thing as a
classical static background potential.

Thus, the background field model is, at best, a crude prescription for
completing incomplete systems.  This may be a sensible procedure if
one is forced to discuss an apparent counterexample that fails to
include all essential parts of a physical object.  But it does not
itself constitute a satisfactory formulation of the Bekenstein bound,
and it cannot be the last word if the bound is to be ascribed any
fundamental significance.

However, the success of the background field model does suggest that
if Bekenstein's bound has a precise formulation, interactions are of
the essence.  Instead of patching up fake counterexamples, it would be
preferable to eliminate incomplete systems from the start.  One would
like to restrict the application of the bound to systems which {\em
confine themselves\/} through interactions, such as a nucleus, an
atom, or a crystal.  This forces us to abandon the simple model of a
finite system as a set of free fields confined by external boundary
conditions.  Instead, we must consider the proper physical description
in terms of interacting fields with no external boundary conditions,
such as QCD, QED, the standard model, or some low energy Lagrangians
obtained from string theory.

Moreover, a statement of the bound with any claim to generality should
not require the specification of a particular macroscopic system
(e.g., a block of iron) before microscopic states are counted.
Instead, it should define the entropy $S$ directly in terms of states
in the Fock space of a relevant quantum field theory.

Therefore we propose that complete systems should be formally
characterized as {\em bound states}, i.e., eigenstates of the full
interacting Hamiltonian which are discrete (have no continuous quantum
numbers).  The entropy $S$ will still be defined as the logarithm of
the number of orthogonal quantum states satisfying macroscopic
conditions specified on the right hand side of the bound.  But we
propose to discard scattering states and take $S$ to count only bound
energy eigenstates.

Of course, any bound state gives rise to a continuous family of
eigenstates related by overall boosts.  In order to mod out by this
trivial continuum, one may fix the total three-momentum to take some
arbitrary but fixed value.  The choice ${\mathbf P}=0$, the rest
frame, has been implicit throughout this paper and in the literature.
Then the energy eigenvalue of each bound state is simply its rest
mass.

Now one may formulate the Bekenstein bound as follows.  Pick upper
limits on the rest mass, $M$, and on the width, $a$.  Define ${\cal
N}(M,a)$ as the number of bound energy eigenstates with
vanishing spatial momentum, eigenvalue $E\leq M$, and support over a
region of width $a$. With 
\begin{equation}
S(M,a)\equiv\log{\cal N}(M,a),
\label{eq-di}
\end{equation}
the bound takes the form
\begin{equation}
S(M,a)\leq\pi Ma.
\label{eq-z4}
\end{equation}

\acknowledgments

I am grateful to M.~Aganagic, N.~Arkani-Hamed, T.~Banks,
J.~Bekenstein, D.~Kabat, D.~Marolf, E.~Martinec, S.~Shenker,
A.~Strominger, L.~Susskind, N.~Toumbas, and especially \'E.~Flanagan
for discussions and comments.  I thank the Harvard physics department,
the Radcliffe Institute, the 2003 Benasque workshop on string theory,
and the Kavli Institute for Theoretical Physics for their hospitality
while this work was completed.

\bibliographystyle{board}
\bibliography{all}
\end{document}